\documentclass[12pt]{article}
\usepackage{epsf}

\begin{document}

\begin{titlepage}
\rightline{hep-ph/0002005}
\rightline{TIFR/TH/00-07}
\vspace{1cm}

\begin{center}
{\Large\bf Understanding Fields Using Strings:}\\[3mm]
{\Large\bf A Review for Particle Physicists}\footnote{Based on an invited 
talk given at the XIII DAE
Symposium on High Energy Physics, Chandigarh, India, December 26-31 1998, to
appear in a special issue of {\it Pramana} Journal of Physics.}\\[6mm]
Sunil Mukhi\\[2mm]
{\it Tata Institute of Fundamental Research}\\
{\it Homi Bhabha Rd, Mumbai 400 005, India}
\end{center}

\vspace{0.5cm}

\begin{abstract}

In addition to being a prime candidate for a fundamental unified
theory of all interactions in nature, string theory provides a natural
setting to understand gauge field theories. This is linked to the
concept of ``D-branes'': extended, solitonic excitations of string
theory which can be studied using techniques of string theory and
which support gauge fields localized along their world-volumes. It
follows that the techniques of string theory can be very useful even
for those particle physicists who are not specifically interested
in unification and/or quantum gravity. In this talk I attempt to
review how strings help us to understand fields. The discussion is
restricted to 3+1 spacetime dimensions.

\end{abstract}
\end{titlepage}

\def\tphi{{\tilde\phi}}
\def\tPsi{{\tilde\Psi}}
\def\tQ{{\tilde Q}}
\def\be{\begin{equation}}
\def\ee{\end{equation}}
\def\bea{\begin{eqnarray}}
\def\eea{\end{eqnarray}}
\def\half{{1\over 2}}
\def\cN{{\cal N}}
\def\cmc{{\cal M}_c}
\def\cmq{{\cal M}_q}
\def\cm{{\cal M}}

\section{Introduction}

A long-standing goal in theoretical high-energy physics is to
understand the dynamics of gauge theory beyond perturbation
theory. This is particularly important for QCD where non-perturbative
effects are responsible for many, if not most, of the physical
behaviour of the theory. While the lattice offers one hope to address
this problem in a very explicit way, it is often the case that a
continuum picture, even qualitative, can be a rich source of insight.

Such a continuum picture --- of confinement and other non-perturbative
effects in gauge theory --- has been conjectured by many outstanding
physicists since the early 1970's. While beautiful and reasonably
convincing physical pictures of QCD emerged from this analysis, it
proved very hard to substantiate much of this thinking by
evidence, even ``theoretical evidence''. 

What is theoretical evidence? In the last five years, we understand
this term much better. Conjectures about strongly-coupled gauge theory
cannot be proved without having a definite and effective computational
procedure in mind, and this is still lacking at present. But there is
a more realistic goal: having formulated a conjecture, one can make a
large list of its consequences, and then hope to isolate, from this
list, a few consequences that can actually be theoretically
tested. This then constitutes a body of theoretical evidence for the
conjecture.

Supersymmetry and string theory have turned out to be the twin planks
on which a large body of theoretical evidence, embodied in ``duality
symmetries'', has accumulated over the last few years. Since not all
high-energy physicists are interested in the goal of string theory (to
unify all four fundamental interactions, including gravity, into a
consistent quantum theory), I have chosen to focus this talk on the
areas in which a more or less conventional particle physicist can gain
insight from string theory. 

Along with string theory, supersymmetry will turn out to be a key
ingredient in our story. Supersymmetry is accepted by most high-energy
physicists as a plausible proposal for what the world is like above a
TeV or so. Even if this proposal turns out not to be correct, string
theory might still be a helpful way to understand the correct field
theory. This is because many of the ``miraculous'' symmetries of
string theory that we will use, might well be present even in the
absence of supersymmetry. It is our knowledge of non-supersymmetric
string theory that is still insufficient to put it to the service of
gauge theory.

This review should be fairly accessible to readers who are not
knowledgeable about string theory. Such readers may, however, wish to
consult Refs.\cite{gsw,polchinski-book} to learn more about the
subject.

\section{Classical SUSY Gauge Theory}

To set the stage for our discussions, it is useful to review the
structure of supersymmetric gauge fields in 4 spacetime
dimensions\cite{wessbagger,mishra}. Supersymmetry requires different
bosonic and fermionic fields to fall into multiplets.

\subsection{Multiplets and Lagrangians}

\noindent \underbar{\it $\cN=1$ Supersymmetry:} With $\cN=1$ 
supersymmetry we have two possible multiplets. The first is a vector
multiplet:
\be
{\rm vector~multiplet}:\qquad A^a_\mu, \lambda^a,\quad a=1,\ldots,{\rm 
dim~G}
\ee
consisting of a gauge field and a Majorana spinor (``gaugino'') in the
adjoint of the gauge group G.

The second multiplet, called the ``chiral multiplet'', contains no gauge
fields but only scalars and fermions:
\be
{\rm chiral~multiplet}:\qquad \phi_I^i, \psi_I^i,\quad
i=1,\ldots,{\rm dim~R},~~ I=1,\ldots,N_f
\ee
where R is a representation of the gauge group G, and $N_f$ denotes
the number of flavours. The scalars in this multiplet are usually
called ``squarks'' since that is what they would be if we were writing 
a supersymmetric version of the standard model.

Together, supersymmetry and gauge symmetry constrain the most general
renormalizable classical action one can write with these fields. The
action is made up of three terms:
\be
S = S_{kinetic} + S_{D-term} + S_{superpotential}
\ee
where $S_{kinetic}$ contains the usual kinetic terms for all the
fields, and
\bea
&S_{D-term} &= \int d^4x\, \sum_{a=1}^{\rm dim~G}\left({\phi_I^i}^\dag
T^a_{ij} \phi^j_I\right)^2 + {\rm fermions}\nonumber\\
&S_{superpotential} &= \int d^4x\, \sum_{i,I} \left| {\partial W\over
\partial \phi^i_I}\right|^2 + {\rm fermions}
\eea
where $W(\phi_i^I)$ is an analytic function of the complex field on
which it depends, and is called the ``superpotential''. Supersymmetry
allows this to be arbitrary, but for renormalizability it should be at 
most cubic in its argument.
\medskip

\noindent \underbar{\it $\cN=2$ Supersymmetry:} With $\cN=2$ 
supersymmetry we again have two possible multiplets. The first is
again called a vector multiplet but its content is different from the
$\cN=1$ vector multiplet:
\be
{\rm vector~multiplet:}\qquad 
A^a_\mu, \Phi^a, \lambda^a, \quad a = 1 \ldots {\rm dim~G} 
\ee
Here $\lambda^a$ is a Dirac gaugino, while $\Phi^a$ is a complex scalar
field in the adjoint. This multiplet is actually the combination of a
vector and a chiral multiplet of $\cN=1$ supersymmetry.

The second multiplet of $\cN=2$ supersymmetry is called a
``hypermultiplet'' and has the following content:
\be
{\rm hypermultiplet:}\qquad Q^i_I, \tQ^i_I, \Psi^i_I,
\tPsi^i_I,\qquad
i = 1 \ldots {\rm dim~R},~~ I = 1 \ldots N_f
\ee
where $Q^i_I$ and $\tQ^i_I$ are complex scalars in the representation
$N_c$ and ${\bar N_c}$ respectively of the gauge group, and $\Psi^i_I,
\tPsi^i_I$ are Weyl fermions in the same representations. Thus the
hypermultiplet is a combination of two chiral multiplets of $\cN=1$
supersymmetry, in conjugate representations.
 
The most general renormalizable action compatible with $\cN=2$
supersymmetry is:
\be
S = S_{\rm kinetic} + S_2 + S_{\rm superpotential}
\ee
where
\bea
S_2 &=& \int d^4x \sum_a |f^{abc} \bar\Phi^b \Phi^c|^2 + {\rm
fermions} \nonumber \\[2mm] 
S_{\rm superpotential} &=& \int d^4x |W'|^2 + {\rm fermions} \nonumber
\\[2mm] 
W(\Phi^a,Q^i_I,\tilde Q^i_I) &=& \tilde Q^i_I \Phi^a T^a_{ij} Q^j_I +
m_{IJ} \tilde Q^i_I Q^i_j
\eea
There is an $SU(2) \times U(1)$ $R$-symmetry under which 
$\lambda^a$ decomposes into a doublet. The squarks $(Q, \tilde Q^+)$
also form an $SU(2)_R$ doublet.
\medskip

\noindent\underbar{\it $\cN=4$ Supersymmetry:} With $\cN=4$ 
supersymmetry there is only a single multiplet, called the vector
multiplet:
\be
{\rm vector~multiplet:}\qquad A^a_\mu, \Phi^a_r, \lambda^a_R,\qquad 
a = 1 \ldots {\rm dim~G},~r = 1 \ldots 6,~R = 1 \cdots 4
\ee
In terms of $\cN=2$ super-multiplets, this is a combination of a
vector multiplet and an adjoint hypermultiplet, while in $\cN=1$
language this is a combination of a vector multiplet and three adjoint 
chiral multiplets. 

With such a high degree of supersymmetry, the action is completely
determined if we allow only renormalizable (dimension 4)
interactions. It takes the form:
\bea
S &=& S_{\rm kinetic} + S_2 \nonumber \\[2mm] 
S_2 &=& \int d^4x \left(\sum^6_{r,s=1} |f^{abc} \Phi^b_r \Phi^c_s|^2 +
{\rm fermions}\right)
\eea
This is the maximally supersymmetric situation if we restrict
ourselves to field theories in 3+1 dimensions without gravity. In
components, there are 16 supersymmetry charges (4 Majorana spinors of 4 
components each). 

\subsection{Classical Parameter Space (``Moduli Space'')}

The parameter space, or ``moduli space'', of a field theory is the
space of degenerate vacuum configurations. This amounts to the space
of energy-minimising vacuum expectation values of various scalar
fields. Classically, this is easy to determine by examining the
Lagrangian and looking for flat directions in field space along which
the potential does not vary. Quantum mechanically, one has to replace
the Lagrangian by the effective Lagrangian incorporating quantum
corrections. 

Without supersymmetry, there is often no moduli space since the
potential will have a unique minimum. Even if we choose a potential
with flat directions, quantum corrections will generically lift this
degeneracy. However, with supersymmetry, the classical moduli space is 
already constrained and moreover, quantum corrections can fail to lift 
degeneracies because of cancellations between fermion and boson loops.
We will denote the classical moduli space by $\cmc$ and the quantum
moduli space by $\cmq$.

The moduli spaces are most constrained when there is the greatest
degree of supersymmetry. Hence in this discussion we start with the
maximally supersymmetric case.
\medskip

\noindent\underbar{\it $\cN=4$ Supersymmetry:} In this case, the
classical moduli space consists of those vacuum expectation values of
the 6 scalars which together minimise the potential energy. The result 
is simple but interesting. We require:
\be
\sum_{r,s} (f^{abc} \phi^b_r \phi^c_s)^2 = 0
\ee
where the scalar fields are understood to represent the
VEV's. Positivity implies
\be
f^{abc} \phi^b_r \phi^c_s = 0 ~{\rm for~all}~a,r,s
\ee
This condition will be satisfied if and only if the VEV's all lie in
the Cartan subalgebra of the gauge group:
\bea
\phi^\alpha_r && ~~ {\rm arbitrary}, ~\alpha
= 1 \cdots ~{\rm rank~G}\nonumber \\
\phi^a_r && = 0, ~a = ({\rm rank~G}) + 1, \ldots, {\rm
dim~G}
\eea
Recall that $r$ takes values from 1 to 6, labelling the 6 scalar
fields in the vector multiplet. 

As a simple example, with gauge group $SU(2)$, we have
\be
\phi^3_r ~~{\rm arbitrary}, ~ \phi^{1,2}_r = 0~~(r=1,\ldots,6)
\ee
It is convenient to label the VEV's by a collection of 6-vectors:
\be
\phi^\alpha_r = v^\alpha_r = (v^\alpha_1,
v^\alpha_2, \ldots , v^\alpha_6) = {\vec v}^\alpha
\ee
Then, the classical moduli space is the space of all 6-vectors ${\vec
v}^\alpha$. However, there are global identifications by the Weyl
group of $G$, a discrete subgroup which must still be imposed as a
gauge symmetry. Thus the true classical moduli space is really the
quotient of the naive one by this group.

The Weyl group of $SU(2)$ is just $Z_2$, while for general $SU(N_c)$ it
is the permutation group $S_{N_c}$. Thus the classical moduli space in 
these cases is:
\bea
SU(2): \cmc &=& {\bf R}^6/Z_2 \nonumber\\
SU(N_c): \cmc &=& {\bf R}^{6(N_c-1)}/S_{N_c}
\eea

Let us consider the $SU(2)$ case in more detail. {\bf R}$^6$ has
coordinates $(v_1, \ldots v_6) = {\vec v}$. The action of the Weyl
group is:
\be
Z_2:\quad {\vec v} \to - {\vec v}
\ee
There is a fixed point of this action at $\vec v = \vec 0$. This is
the point where $SU(2)$ gauge symmetry is restored, since the adjoint
scalar VEV's all vanish. Elsewhere, $\phi^3_r = v_r \neq 0$ 
breaks $SU(2)$ to $U(1)$.

Geometrically, a fixed point of the quotienting group corresponds to a
singularity of the space. The space becomes an {\it orbifold}, so
while it is flat everywhere else, it has infinite curvature at the
origin.

Far away from the origin ($\vec v \neq \vec 0$), the off-diagonal
$SU(2)$ gauge particles, which we may denote $W^\pm$, are massive,
with a mass $g_{YM} |\vec v|$. As $\vec v \rightarrow \vec 0$, these
gauge particles become massless. We see that singularities of the
moduli space $\cm$ are associated to the presence of new massless
particles in the spectrum.

For $SU(N_c)$, at a generic point of $\cmc$ we have the
symmetry-breaking pattern:
\[
SU(N_c) \to (U(1))^{N_c-1}
\]
Note that the Cartan subgroup $(U(1))^{N_c-1}$ of $SU(N_c)$ can never
be broken by the VEV of an adjoint scalar (since adjoint scalars are
uncharged under this subgroup). Hence at such generic points we always 
have a number (${\rm rank~G}$) of massless photons, and the theory is in 
the Coulomb phase.

However, there are special points where the breaking pattern is
different:
\bea
SU(N_c) &\to& SU(2) \times (U(1))^{N_c-2} \nonumber \\ [2mm]
&\to& SU(3) \times (U(1))^{N_c-3} \nonumber \\ [2mm]
&\to& SU(2) \times SU(3) \times U(1) \times \cdots
\eea
and so on. All such points have ``enhanced nonabelian symmetry'',
hence extra massless particles. These points are fixed under the
action of some element of $S_{N_c}$, hence they are singularities of
the moduli space.

In addition to the above moduli space, there is the 
parameter space for the gauge coupling $g_{YM}$ and the
$\theta$-angle, which combine into a complex parameter:
\be
\tau_{YM} = {\theta \over 2\pi} + {4\pi i \over g^2_{YM}}
\label{yangmills}
\ee
In field theory these parameters are fixed by hand and are quite
distinct from VEV's of scalar fields. However, in string theory they
arise as VEV's of some appropriate scalar fields, hence string
theorists usually consider this parameter space to be part of the
moduli space.
\medskip

\noindent \underbar{\it $\cN=2$ Supersymmetry:} In this case, there
are other phases besides the Coulomb phase. Thus the classical moduli
space $\cmc$ splits into branches. One branch is characterised by the
following:
\bea
Q^i_I &=& {\tilde Q^i_I} = 0\nonumber\\ 
\phi^\alpha &=& v^\alpha, \alpha = 1, \ldots, {\rm rank~G}
\eea
Note that with $\cN=2$ supersymmetry, the field $\phi^\alpha$ and its
VEV $v^\alpha$ are {\it complex} numbers.

The above equation defines the ``Coulomb branch'', on which as before, 
the generic breaking pattern is:
\be
SU(N_c) \to (U(1))^{N_c-1}
\ee
In particular, for $SU(2)$ we have the Coulomb branch:
\be
\cmc^{Coulomb} = {\bf R}^2/Z_2
\ee
where $v = v^3$ is the (complex) coordinate on ${\bf R}^2$.

At generic points of $\cmc^{Coulomb}$ we cannot give a VEV to
$Q^i_I,{\tilde Q}^i_I$, since their couplings to the adjoint scalars
would increase the potential energy. But it is not hard to see that if
$v^\alpha$ takes some special values, then we can turn on VEV's for
$Q^i_I, {\tilde Q}^i_I$ at no cost in energy. Since the
hypermultiplets are usually in the fundamental representation, they
are charged under the Cartan subgroup of the gauge group. Hence such
VEV's break even $U(1)$ factors. This branch of the moduli space is
therefore called the Higgs branch.
\medskip

\noindent \underbar{\it $\cN=1$ Supersymmetry:} In this case the
vector multiplet contains no scalars, hence there is no moduli space
unless we couple some chiral (matter) multiplets. With matter, we have
to minimize 
\[
S_{D-term} + S_{superpotential}
\]
The result for $\cmc$ depends on the details of the fields,
representations and choice of superpotential. Not much can be said
about it without going into a detailed classification of cases.

We see that the classical moduli space $\cmc$ is relatively simple for
$\cN=4$ and consists of a Coulomb phase, while for $\cN=2$
supersymmetry, it consists of intersecting Coulomb and Higgs
branches. With $\cN=1$ supersymmetry, the moduli space depends largely 
on one's choice of field content and superpotential in the theory. 

\section{Quantum SUSY Gauge Theory}

We now turn to the question of how quantum corrections modify the
classical moduli space of a supersymmetric gauge theory. In general,
the quantum effective action will be different from the classical one
and will incorporate non-renormalizable terms, including more general
kinetic terms than the usual ones. 
\medskip

\noindent \underbar{\it $\cN=4$ Supersymmetry:} Because of the high 
degree of supersymmetry, the quantum moduli space $\cmq$ is identical
to the classical one $\cmc$. At the origin of $\cmq$, the theory has
unbroken $SU(N_c)$ gauge symmetry and vanishing $\beta$-function.
Thus, it is a conformal field theory (CFT). Note that as a consequence
there is no asymptotic freedom, and hence also no confinement, in this
theory. Away from the origin, conformal invariance is broken by the
scalar VEV and we have massive theory coupled to $U(1)$ gauge fields. 
\medskip

\noindent \underbar{\it $\cN=2$ Supersymmetry:} Consider $SU(2)$ gauge 
theory with no hypermultiplets. It was shown non-perturbatively, by
Seiberg and Witten\cite{seiwit-one,seiwit-two}, that the structure of
$\cmq$ is rather different from that of $\cmc$. Whereas in $\cmc$ the
Coulomb branch is singular at the origin and $SU(2)$ gauge symmetry is
restored there, in $\cmq$ the Coulomb branch has no singularity at the
origin. Moreover, in this theory $SU(2)$ gauge symmetry is {\it never}
restored at any point of the moduli space!

Instead, it is found that there are two other singular points in
$\cmq$. At these points, some particles do become massless -- but not
the gauge bosons. The massless particles at these points are monopoles
and dyons. It becomes useful to make an electric-magnetic duality
transformation near these points and study the magnetic theory instead. 

This $N=2$ theory has a nontrivial $\beta$-function and is
asymptotically free, so the coupling $\tau = {\theta \over
2\pi}+{4\pi i \over g^2_{YM}}$ depends on the scale. This coupling was
shown to vary complex analytically (``holomorphically'') as a function
of the complex VEV $\phi^3 = v$: so we can write $\tau =
\tau(v)$. This dependence is known exactly as a certain non-trivial
``fibre bundle''. Since $\tau$ is valued in the upper half plane, it
can naturally be interpreted as the ``shape'' parameter (technically,
``complex structure parameter'') of a torus, thus the moduli space
looks like a torus varying over a plane.

The above holds for $SU(2)$ gauge group and $N_f=0$ (no
matter). Analogous exact results for $\cN=2$ supersymmetry are also
known for $SU(N_c)$ gauge groups and for $N_f \leq 2N_c$ flavours, for
which the theories are always asymptotically free. For $N_f = 2N_c$
these theories are finite (the $\beta$-function vanishes) and hence
they are conformal field theories. For $N_f > N_c$ the
$\beta$-function is positive and the theory becomes ill-defined.

Note that the interesting results about quantum corrections always
concern the Coulomb branch. The Higgs branch is protected from quantum
corrections. 
\medskip

\noindent \underbar{\it $\cN=1$ Supersymmetry:} A complex 
array of results have been found for the quantum moduli space of $N=1$
supersymmetric gauge theories. However, just as the classical moduli
space in this case depends on the detailed choice of matter fields,
representations and couplings, the structure of $\cmq$ too will depend
on these choices. The interested reader is referred to appropriate
review articles on this topic, such as Ref.\cite{intsei}.

\section{D-branes and $\cN=4$ SUSY} 

In this section we show how supersymmetric gauge theories in 3+1
spacetime dimensions naturally arise as a subsector of superstring
theory. For a more detailed review of the relevant material on
D-branes, see Ref.\cite{polchinski-book}.

Introducing fundamental extended objects such as strings leads to a
variety of interesting new physical consequences. For one thing,
closed string excitations produce gravity, so string theories are
theories of quantum gravity. But we will be more interested in the
sector of string theory that contains open strings.

Open strings have a pair of ends. This requires the specification of
boundary conditions at the endpoints. While it is most natural to
allow these to lie anywhere in space, one can consistently choose to
restrict the endpoints onto a $p$-dimensional spatial hypersurface in
the 9-dimensional space where strings propagate. In fact, one can show
that such choices must necessarily be consistent: starting with
unconstrained endpoints and applying known symmetries of string theory,
we end up with endpoints constrained on a hypersurface.

What is the physical interpretation of these constrained endpoints?
They define a spatial region on which the strings can end. Suppose we
choose $p=0$ and constrain our open strings to end on a fixed point in
space. Then, that point breaks translation invariance exactly as an
elementary particle would do. (For example, applying a Lorentz boost
to the theory would cause the point to start moving with a fixed
velocity). Fluctuations of the string give rise to motions and
oscillations of this fixed endpoint. Hence in all respects this string 
endpoint can be treated as a particle with a definite mass. Because
constrained endpoints satisfy Dirichlet boundary conditions, we call
the associated particle a ``D-particle''.

D-particles can also be understood as solitonic excitations in the
string theory. Hence we have two different mental pictures of the same
object: as a soliton, and as a string endpoint. Now suppose we choose
$p=2$ instead of 0. Then the string endpoint sweeps out a
2-dimensional space. The associated object looks like a
membrane. Indeed, it is called a ``D-brane''. It too has a
complementary description as an extended solitonic excitation in
string theory, much like the cosmic strings and domain walls that can
be found as classical solutions of more physically relevant field
theories. For arbitrary $p$, we say that the string endpoint describes 
a D$p$-brane.

For suitable values of $p$, D$p$-branes are stable objects in type II
superstring theory. They are charged under some generalised gauge
field and hence, in the solitonic picture, they correspond to stable
solitons.

A key property of open superstrings is that their lowest excitations
are massless gauge fields. These gauge fields propagate only on the
locus where the endpoints are free to move, namely on the
D$p$-brane. Thus, the low-energy field theory coming from the dynamics
of open strings is a gauge theory in $p+1$ spacetime dimensions. This
is the central observation that links string theory and gauge field
theory. For our purposes we will select the value $p=3$, so we intend
to realise the supersymmetric field theories discussed in the
preceding sections as modes of open strings ending on D3-branes. The
underlying string theory which has stable D3-branes is called type IIB 
string theory. 

Because of supersymmetry, the gauge fields arising from open string
endpoints lie in supermultiplets containing scalars and fermions. The
basic D3-brane of type IIB string theory can be shown to inherit
$\cN=4$ supersymmetry from the underlying spacetime supersymmetry of
the 10-dimensional string theory. Hence the theory on the worldvolume
of a single D3-brane is an $\cN=4$ supersymmetric gauge theory. A
single D3-brane gives rise to Abelian gauge theory. We will argue
below that to get higher gauge groups one must stack several identical
D3-branes together. We will also see that lower supersymmetry can be
obtained by combining D3-branes with other D-branes and
``orientifolds''.

Before doing this, let us note one amusing fact. A soliton has
``collective coordinates'' for the symmetries that it breaks. In
particular, extended solitons (branes) break translational invariance
in the directions transverse to their own world-volume.  For example,
suppose a D3-brane is arranged to lie along $(x^1, x^2, x^3)$. It
breaks the remaining 6 translational symmetries in the $9+1$
dimensional string theory, along $(x^4, x^5, \cdots , x^9)$. So it
should have 6 massless scalar fields on its world-volume. And it does, 
because $\cN=4$ supersymmetry requires precisely 6 scalar fields in a
vector multiplet!

We see that the 6 scalar fields in the $\cN=4$ supersymmetry
multiplet, whose presence was deduced from the supersymmetry algebra
long before superstrings and D-branes were understood, are most
naturally interpreted as translational collective coordinates of a
D3-brane. Moreover, the $SO(6)$ R-symmetry comes from transverse
rotational invariance: the 10-dimensional Lorentz group $SO(9,1)$ is
broken by the D3-brane into $SO(3,1)\times SO(6)$. Thus R-symmetry (a
key property of field theories with extended supersymmetry) gets
re-interpreted as a spacetime symmetry.

Now consider two parallel D3-branes (Fig.1). Both are aligned along
$(x^1,x^2,x^3)$ but they can be at arbitrary locations in the other
six directions. We let the vector ${\vec v}$ denote the relative
location of one brane with respect to the other along these
directions.

\begin{figure}[htbp]
\epsfxsize=8cm
\centerline{\epsfbox{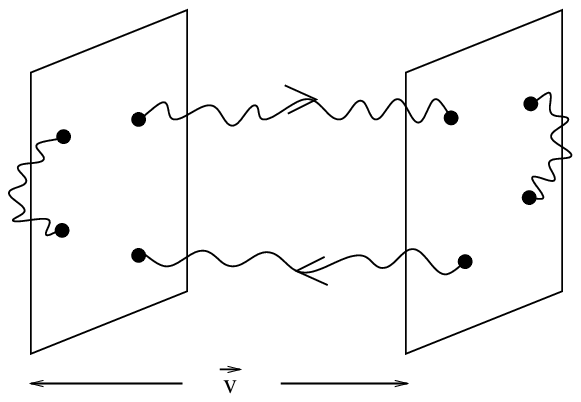}}
\caption{Two parallel D3-branes.}
\label{fig:one}
\end{figure}

From the previous discussion we should expect that together, these
D3-branes support a $U(1) \times U(1)$ $\cN=4$ supersymmetric gauge
theory. The two vector multiplets arise from open strings having both
ends on the first brane or both ends on the second brane. But now we
also have two more types of open strings: those beginning on the first
brane and ending on the second, and those beginning on the second
brane and ending on the first.  The corresponding states are charged
as $(1,-1)$ and $(-1,1)$ under $U(1)\times U(1)$. Under the diagonal
$U(1)$ they are neutral. With respect to the other $U(1)$, they have
exactly the charges of massive W-bosons! In fact their mass is
\be
\label{masscharge}
m_W \sim T |\vec v|
\ee
where $T$ is the string tension. In suitable units, this is related to 
the Yang-Mills coupling constant for the D3-brane gauge theory by
$T\sim {1\over g_{YM}^2}$. (Note that $\vec v$ in this section is a
distance, while in the previous sections it was the VEV of a scalar
field. The translation between these two involves a change of units
and some rescaling.)

Particles obeying a mass-charge relationship like the one above
correspond to quantum states in the gauge theory that do not break all
the underlying supersymmetry (as a generic state would do) but
preserve a fraction of supersymmetry. Such states are known as ``BPS
states'', and the corresponding particles are necessarily stable by
virtue of the supersymmetry that they preserve.

Thus, two parallel D3-branes realize the Coulomb branch of $N=4$
$SU(2)$ gauge theory (apart from a decoupled centre-of-mass $U(1)$)
\cite{witten-bound}. When the parallel D3-branes coincide, the stretched
open strings shrink to zero length, and $|\vec v| = 0$. There, $SU(2)$ is
restored. This is the origin of the Coulomb branch.

Since D-branes are indistinguishable objects, the parameter space is
${\bf R}^6/Z_2$, as we predicted from purely field-theoretic
considerations. Thus we see that in string theory, the Weyl group
factor in the gauge group comes from D-brane statistics!

For $N_c$ parallel, separated D3-branes we have the following picture.
The total number of stretched strings between pairs of D3-branes is
$N_c(N_c-1)$.  Add $N_c$ strings that begin and end on the same brane,
and we end up with $N_c^2$ fields altogether. This is the dimension of
the group $U(N_c)\sim SU(N_c)\times U(1)$.  So, $N_c$ parallel
D3-branes describe the moduli space of $U(N_c) \sim SU(N_c) \times
U(1)$ $\cN=4$ supersymmetric gauge theories (Fig.2).

\begin{figure}[htbp]
\epsfxsize=8cm
\centerline{\epsfbox{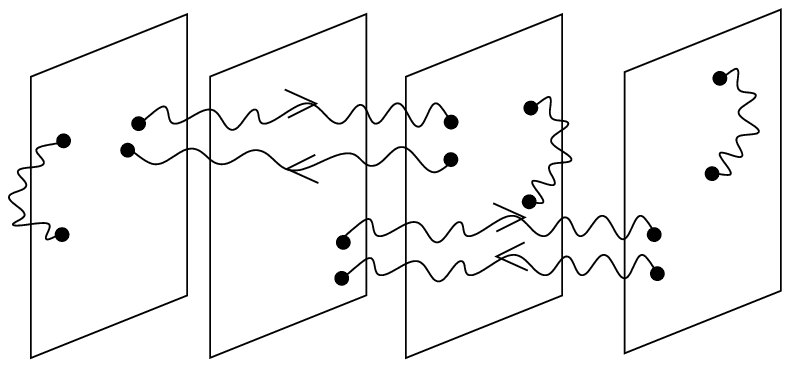}}
\caption{$N_c$ parallel D3-branes.}
\label{fig:two}
\end{figure}

We have already identified some stable BPS states $(W^\pm$ bosons) in
these theories. These carry electric charge under the $U(1)$
factors. Now let us use string duality to extract more information. The
type IIB string in 10 dimensions has a pair of massless scalar
particles: the dilaton $\varphi$, and the axion $\tilde\varphi$. These
appear naturally in the complex combination
\be
\tau_s = {\tilde\varphi \over 2\pi} + 4\pi i e^{-\varphi} =
{\tilde\varphi \over 2\pi} + {4\pi i \over g_s}
\ee
We have used the fact, well-known to string theorists, that the string
coupling is determined by the expectation value of the dilaton field:
$g_s = e^\varphi$.

Since the modes propagating on the D3-brane are excitations of open
strings, they ``inherit'' this coupling. In fact, the complex
combination $\tau_{YM}$ of Yang-Mills coupling and theta-angle which
we encountered in Eq.~(\ref{yangmills}) is equal to the complex
combination $\tau_s$ above:
\be
\tau_{YM} = {\varphi \over 2\pi} + {4\pi i \over g^2_{YM}} = \tau_s
= {\tilde\varphi \over 2\pi} + {4\pi i \over g_s}
\ee
Hence, in particular, $g^2_{YM} = g_s$. 

Now, it is believed that type IIB string theory has a group of duality
symmetries, SL(2,{\bf Z}), under which
\be
\tau_s \to {a\tau_s + b \over c \tau_s + d},\qquad\pmatrix{a & b \cr c & d}
~\epsilon~~ {\rm SL(2,{\bf Z})}
\ee
This group of transformations includes, as a special case, a simple
integer shift of the axion,
\[
{\tilde\varphi}\to {\tilde\varphi}+1
\]
which tells us that it is an angle-valued field. It also includes the
more nontrivial strong-weak coupling duality (``S-duality'')
\[
\tau_s \to -{1\over \tau_s}
\]
which for zero axion acts as $g_s\to 1/g_s$, inverting strong
and weak coupling in the string theory.  

Under S-duality, we know how all the massless fields of type IIB
string transform. Since D-branes carry charges under specific massless
fields, this also tells us how the branes transform. In particular one
finds that S-duality converts the fundamental type II string into a
D1-brane or ``D-string'', but leaves the D3-brane invariant.

It follows that $\cN=4$ supersymmetric gauge theory must have a
symmetry under
\[
\tau_{YM} \to {a \tau_{YM} + b \over c \tau_{YM} + d}, 
\]
For vanishing $\theta$-angle, this includes a transformation $g_{YM}
\to - {1 \over g_{YM}}$, which interchanges a weakly coupled gauge
theory with a strongly coupled one. 

In addition, we saw that this duality acts on the string theory to
interchange a fundamental type IIB string with a D-string. But we know 
that the end-point of a fundamental string when it terminates on a
brane behaves like an electrically charged particle of the brane
worldvolume theory. It is also known that the endpoint of a D-string
when it terminates on a brane, behaves like a magnetically charged
particle \cite{strominger,tseytlin}. Thus when acting on a
D3-brane, S-duality must interchange electric with magnetic fields.

Thus, stringy S-duality implies strong-weak, electric-magnetic duality
of $\cN=4$ supersymmetric gauge theory. This in turn implies the
existence of a definite spectrum of monopoles and dyons as a
consequence of the existence of electrically charged W-bosons, which
can be identified as perturbative states. While this result was
originally argued from field-theoretic considerations
\cite{sen-sdual}, this way of understanding it through string theory
is very powerful and conceptually illuminating. (It is not as
rigorous, though, since the string duality that we invoke remains a
conjecture, which is in some ways harder to prove or justify than the
field-theoretic duality).

Here we have seen perhaps the simplest example wherein, by realising a
field theory in terms of worldvolume excitations on a brane, one can
derive properties of this field theory using known (or conjectured)
properties of the underlying string theory. These results are
nonperturbative, since the duality acts non-perturbatively.

For general gauge groups $SU(N_c)$, one re-discovers in this way a
rich and complex spectrum of BPS monopoles and dyons, which field
theorists had been slowly discovering over the last two decades.

For $N_c \ge 3$, $SU(N_c)$ gauge theory also admits exotic BPS dyons
whose existence had been conjectured (but not demonstrated) by field
theorists. Such dyons have electric and magnetic charge vectors that
are not proportional. String theory can be used to show that they must
exist.  One starts with the fact that type IIB string theory admits
BPS junctions where a fundamental string meets a D-string and a
bound-state of the two comes out from the junction point (Fig.3). More
general junctions also exist.

\begin{figure}[htbp]
\epsfxsize=8cm
\centerline{\epsfbox{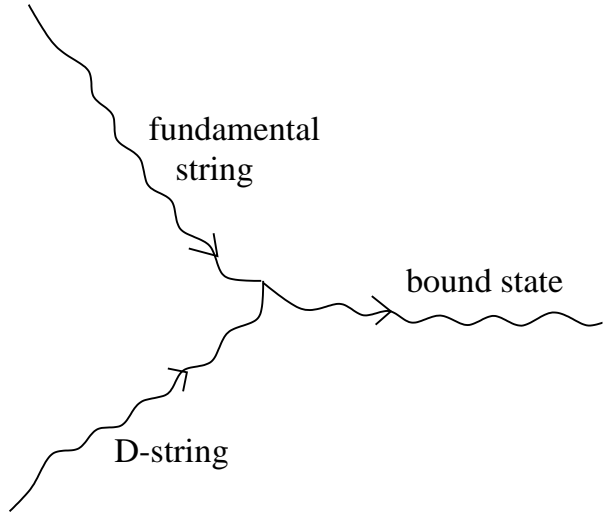}}
\caption{A three-string junction.}
\label{fig:three}
\end{figure}

Following the observation that string junctions are stable, BPS
objects, \cite{dasm-junct}, it was argued \cite{bergman} that an
exotic dyon is obtained by suspending such a string junction between
D3-branes. Such dyons exist in all $SU(N_c)$ $\cN=4$ supersymmetric
theories, for $N_c \geq 3$, and are stable.

With this impetus, field theorists began to generate the appropriate
classical solutions for such field-theoretic solitons, and quite a lot
is known about them by now.

\section{Brane Probes and $\cN=2$}

Type IIB symmetric is invariant under orientation-reversal of the
closed string. This symmetry, denoted $\Omega$, generates a $Z_2$
group and has a definite action on the fields of the theory (and
therefore, as we have seen, also on the branes). Let us compactify IIB 
string theory on a 2-torus, with coordinates $(x^8,x^9)$, and take the
quotient by the symmetry $\Omega\,{\cal I}_{89}$ where ${\cal
I}_{89}$ denotes reflection of the two toroidal directions:
\[
{\cal I}: (x^8, x^9) \rightarrow (-x^8, -x^9) 
\]
As we might expect, $\Omega$ creates unoriented closed strings out of
oriented ones, and ${\cal I}_{89}$ makes the two toroidal space
dimensions into the orbifold ${T^2/Z_2}$ (details about orientifolds
can be found in Ref.\cite{polchinski-book}).
 
The reflection symmetry has 4 fixed points on $T^2$. Let us focus on
one of them, say the one at the origin. This is a point on the
2-torus, but it is independent of the other 7 spatial directions and
is therefore a 7-dimensional hyperplane that extends along those
directions. We call it an ``orientifold 7-plane''.

This object is like a mirror: the spatial regions on the two opposite
sides of it get identified. If we bring a D3-brane near it, we get new
light states coming from open strings joining the D3-brane to its
mirror image (Fig.4). These become massless precisely when the
D3-brane meets the orientifold 7-plane. This leads to two effects. The
7-plane breaks the supersymmetry on the D3-brane (which was originally
$\cN=4$) down to $\cN=2$. The other effect is that out of four open
string sectors on a pair of D3-branes, one is projected out, leading
to an $SU(2)$ gauge group rather than $U(2)$.

\begin{figure}[htbp]
\epsfxsize=8cm
\centerline{\epsfbox{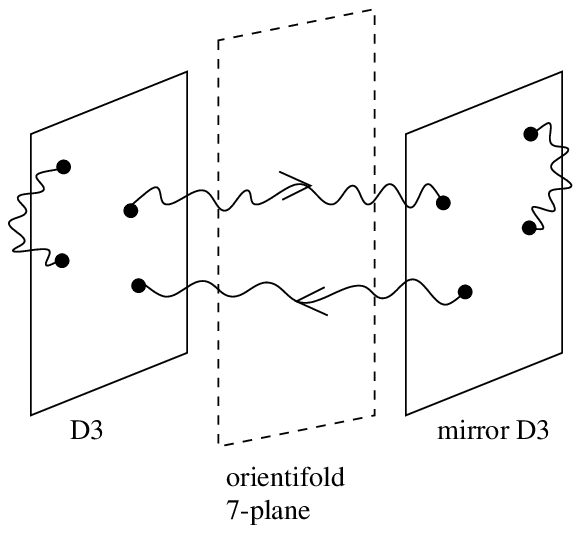}}
\caption{D3-branes at an orientifold 7-plane.}
\label{fig:four}
\end{figure}

The result is pure $\cN=2$ supersymmetric gauge theory with $SU(2)$
gauge group. This means that the moduli space of that theory must be
given by the geometric space encountered by the D3-brane. In fact, we
have recovered the classical moduli space $R^2/Z_2$ of this
theory!

What about quantum effects? In the presence of this orientifold plane,
the type IIB theory becomes a ``type-I'' string theory with reduced
supersymmetry. It was shown \cite{sen-f}, following the construction of
``F-theory'' \cite{vafa-f}, that quantum effects split the orientifold
7-plane into two dynamical 7-branes.

These 7-branes do not allow a type IIB string to end on them. So there
are no massless ``W-bosons'' when the D3-brane touches them. However,
they allow dyonic $(p,q)$ strings (bound states of $p$ fundamental
strings and $q$ D-strings) to end on them. Since the end point of a
fundamental string on a D3-brane is an electric charge, and the
endpoint of a D-string on a D3-brane is a magnetic charge, it must be
true that the endpoint of a $(p,q)$ string is a dyon of electric
charge $p$ and magnetic charge $q$. Hence when the D3-brane touches
either of the D7-branes, we get corresponding massless $(p,q)$ dyons.

We have recovered an essential part of the Seiberg-Witten picture. The
origin of the Coulomb branch has split into two singularities where
dyons become massless. There is no point where W-bosons become
massless.  

To complete the picture, we use the existence of ``F-theory''
\cite{vafa-f}, which is a novel way of compactifying the type IIB 
string where its coupling $\tau_s$ is allowed to vary over the compact
manifold.  Since the D3-brane inherits this coupling, the gauge
coupling $\tau_{YM}$ too varies over the $v$-plane (where $v = x^8 +
i x^9)$) exactly as predicted by Seiberg and Witten. The
Seiberg-Witten torus, which was a mathematical artifact in their
solution, is realised geometrically: it turns out to be the torus
whose shape is parametrised by $\tilde\varphi$ (the axion) and
$\varphi$ (the dilaton).

We can also introduce D7-branes parallel to the orientifold plane,
this gives rise to (massive) hypermultiplets coupled to the pure
$\cN=2$ gauge theory. In this way one recovers the more general
Seiberg-Witten theories incorporating $\cN=2$ matter multiplets, and
the Higgs branch appears as well.

One can use this stringy setup to predict new field-theoretic
phenomena. The usual Seiberg-Witten theories have a maximal flavour
symmetry group $SO(8)$, which is realised in the case of four massless
flavours. However, it was argued \cite{dasm-f} that some
configurations of 7-branes give rise to gauge theories on the 3-brane
with $E_6, E_7, E_8$ global symmetry. This phenomenon (unlike the
familiar $SO(8)$ case) cannot occur at weak coupling. It is a new
non-perturbative field-theoretic effect predicted by string theory.

\section{Large-$N_c$ Gauge Theories and the AdS/CFT Correspondence}

D3-branes have some features that we have not yet
explored. Complementary to their description as D-objects (loci of
open-string endpoints), they can also be understood as solitonic
classical solutions of type IIB string theory -- more specifically, of
its low-energy limit, type IIB supergravity. Hence there is a
spacetime metric describing the gravitational field around a
collection of $N_c$ D3-branes:
\bea
ds^2 &=& f(r)^{-\half} (- dt^2 + (dx^1)^2 + (dx^2)^2 + (dx^3)^2)
\nonumber \\ [2mm]
&& + f(r)^{\half} ((dx^4)^2 + \cdots + (dx^9)^2)
\eea
where
\be
f(r) = 1 + {R^4 \over r^4}, \qquad r = \left( (x^4)^2 + \ldots +
(x^9)^2\right)^{\half}
\ee
and 
\be
\label{Rdef}
R \sim (g_s (\alpha')^2 N_c)^{{1\over 4}}
\ee
This metric describes a massive object localised along three spatial
directions. Some generalised gauge fields of the low-energy
supergravity theory must also be excited to make this a genuine
classical solution. As a result, the solution describes a charged
object. In fact, it is supersymmetric (BPS), and has a mass-charge
relationship analogous to that in Eq.(\ref{masscharge}), except that
mass is replaced by mass per unit 3-volume or ``brane tension''.

Something remarkable happens in the limit of large $R$ (which, from
Eq.(\ref{Rdef}) is the same as large $g_s N_c = g^2_{YM} N_c$). From
the form of $f(r)$ above, this limit is equivalent to the
``near-horizon'' limit $r \ll R$ in which we probe the metric very close
to the brane. In this limit, we can make the replacement
\be
f(r) = 1 + {R^4 \over r^4}~ \to~ {R^4 \over r^4}
\ee
and as a result the spacetime metric around $N_c$ D3-branes becomes:
\bea
ds^2 &=& {r^2 \over R^2} (-dt^2 + (dx^1)^2 + (dx^2)^2 + (dx^3)^2)
+ {R^2 \over r^2} (dr^2 + r^2 (d \Omega_5)^2) \nonumber \\ [2mm]
&=& \left\{{r^2 \over R^2}(-dt^2 + (dx^1)^2 + (dx^2)^2 + (dx^3)^2) + R^2
{dr^2 \over r^2}\right\} + R^2 (d \Omega_5)^2
\eea
The factor in large braces is the metric of a $(4+1)$-dimensional
space-time called ``anti-deSitter'', and denoted $AdS_5$, while the
last term is the metric of a 5-sphere. Thus we have shown that the
near-horizon metric of $N_c$ D3-branes is the space-time $AdS_5 \times
S^5$.

As $N_c$ grows, the near-horizon region expands. In the limit of
infinite $N_c$, the entire spacetime (not just near the branes) is
$AdS_5 \times S^5$. Based on these facts, Maldacena \cite{maldacena}
made a novel conjecture. According to this, the following two
descriptions of D3-branes for large $N_c$ are equivalent:

{(i)} The description as the limit of $\cN=4$ supersymmetric gauge
theory as $(g^2_{YM} N_c)$ becomes large,

{(ii)} The description as the nontrivial spacetime background $AdS_5
\times S^5$ of type IIB string theory.

This is a duality between on one hand a gauge theory, and on the other
hand a theory of gravity and strings. It is remarkable how the
symmetries of the problem match up in the two descriptions. In the
gravity description, we have the symmetry groups $SO(4,2)$ and
$SO(6)$, making up the isometries of the maximally symmetric spaces
$AdS_5$ and $S^5$ respectively. In the gauge theory description,
$SO(4,2)$ is realised as the conformal symmetry group of
$3+1$-dimensional gauge theory, which includes the Poincare group. On
the other hand, $SO(6)\sim SU(4)$ is the R-symmetry group of $\cN=4$
supersymmetric Yang-Mills theory.

If we are only interested in the leading behaviour in the limit of
large $g^2_{YM} N_c$, we can really ignore string theory in favour of
its low energy limit, type IIB supergravity. This is because the
massive stringy modes decouple in this limit. 

Precise prescriptions have been given \cite{witten-ads,gkp} to relate
correlation functions in $\cN=4$ gauge theory to computations in
supergravity. This opens up the possibility of solving the quantum
gauge theory completely in the large-$N_c$ limit just using the
classical Lagrangian of supergravity!

Some of the remarkable results obtained in this direction concern the
computation of expectation values of Wilson loops 
\cite{maldacena-wl,rey-yee}, properties of baryons and domain walls 
\cite{witten-bar}, and thermal properties and phase transitions in 
gauge theory \cite{witten-therm}. The correspondence was also extended
to the case of lower supersymmetry: $\cN=2$, $\cN=1$, and even $\cN=0$
(no supersymmetry) \cite{kachsil,lnv,klebwit}.

An interesting example of lower supersymmetry is a case with $\cN=1$
supersymmetry in four dimensions. This arises by placing $N_c$
D3-branes at the singular tip of a singular noncompact manifold called
a ``conifold''. One finds in this case an interesting $\cN=1$
supersymmetric field theory on the D3-branes, which exhibits a
nontrivial flow in the infrared to a superconformal field theory
\cite{klebwit}. A dual brane description of this was found
\cite{uranga,dasm-conif,dasm-frac} which leads to a
description of the field theory and its symmetries using strongly
coupled string theory or ``M-theory''.

\section{Conclusions}

String theory has found a new role: to help in ``solving'' gauge
theories non-perturbatively. Such solutions range from a qualitative
understanding of the theories, including their symmetries, to a
detailed description of the moduli space in the same sense that
Seiberg and Witten initially achieved using only field theoretic
techniques.

Due to a shortage of time, I could not discuss a fascinating approach
to realising field theories in terms of intersecting branes, the
so-called ``brane constructions'' \cite{hanwit,elgivkut}. These
provide much more general examples of the utility of string theory in
understanding quantum field theory.

Though such constructions exist for various different amounts of
supersymmetry upto the maximal case of $\cN=4$, it remains true that
at present our understanding is best for the most highly
supersymmetric, and hence less interesting, gauge theories. It is
important to improve our understanding of theories with $\cN=1$
supersymmetry, which is the amount of supersymmetry in the MSSM (such
theories are dynamically quite similar to non-supersymmetric
theories). Some partial progress has also been made towards directly
studying non-supersymmetric gauge theories using string theory. The
day may not be far off when the Standard Model will be most easily
understood by representing it as a sector of string theory.

\end{document}